\documentclass[letterpaper]{sig-alternate-2013}

\usepackage[
  pass,
]{geometry}

\usepackage{graphicx}
\usepackage{subfig}
\usepackage{balance}
\usepackage{multirow} 

\permission{\copyright~2017 International World Wide Web Conference Committee (IW3C2), published under Creative Commons CC BY 4.0 License.}
\conferenceinfo{WWW'17 Companion,}{April 3--7, 2017, Perth, Australia.}
\copyrightetc{ACM \the\acmcopyr}
\crdata{978-1-4503-4914-7/17/04. \\
http://dx.doi.org/10.1145/3041021.3051153 \\
\includegraphics{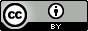}
}
 
\begin{document}

\title{Extracting Implicit Social Relation for Social Recommendation Techniques in User Rating Prediction}
%
%
%
%
%

\numberofauthors{6} 
%
\author{
%
%
\alignauthor
Seyed Mohammad Taheri\\
       \affaddr{Department of Computer Engineering, Sharif University of Technology}\\
       \email{mtaheri@ce.sharif.edu}
\alignauthor
Hamidreza Mahyar\\
       \affaddr{Department of Computer Engineering, Sharif University of Technology}\\
       \email{hmahyar@ce.sharif.edu}
\alignauthor Mohammad Firouzi\\
       \affaddr{Department of Computer Engineering, Sharif University of Technology}\\
       \email{mfirouzi@ce.sharif.edu}
\and  
\alignauthor Elahe Ghalebi K.\\
       \affaddr{Department of Computer Engineering, Vienna University of Technology}\\
       \email{eghalebi@cps.tuwien.ac.at}
\alignauthor Radu Grosu\\
       \affaddr{Department of Computer Engineering, Vienna University of Technology}\\
       \email{radu.grosu@tuwien.ac.at}
\alignauthor Ali Movaghar\\
       \affaddr{Department of Computer Engineering, Sharif University of Technology}\\
       \email{movaghar@sharif.edu}
}
\date{20 February 2017}

\maketitle
\begin{abstract}
Recommendation plays an increasingly important role in our daily lives. Recommender systems automatically suggest items to users that might be interesting for them. Recent studies illustrate that incorporating social trust in Matrix Factorization methods demonstrably improves accuracy of rating prediction. Such approaches mainly use the trust scores explicitly expressed by users. However, it is often challenging to have users provide explicit trust scores of each other. There exist quite a few works, which propose Trust Metrics to compute and predict trust scores between users based on their interactions. In this paper, first we present how social relation can be extracted from users' ratings to items by describing Hellinger distance between users in recommender systems. Then, we propose to incorporate the predicted trust scores into social matrix factorization models. By analyzing social relation extraction from three well-known real-world datasets, which both: trust and recommendation data available, we conclude that using the implicit social relation in social recommendation techniques has almost the same performance compared to the actual trust scores explicitly expressed by users. Hence, we build our method, called Hell-TrustSVD, on top of the state-of-the-art social recommendation technique to incorporate both the extracted implicit social relations and ratings given by users on the prediction of items for an active user. To the best of our knowledge, this is the first work to extend TrustSVD with extracted social trust information. The experimental results support the idea of employing implicit trust into matrix factorization whenever explicit trust is not available, can perform much better than the state-of-the-art approaches in user rating prediction.
\end{abstract}

\\
\section{Introduction}

The explosive growth of available information brings forth the \lq\lq information overload \rq\rq problem. Recommender Systems (RS) are information filtering systems that deal with the problem of information overload by filtering vital information fragment out of large amount of dynamically generated information according to user's preferences, interests, and observed behavior about items \cite{isinkaye2015recommendation}. Recommender systems help users with item selection and purchasing decisions based on users' tastes and preferences using a variety of information gathering techniques \cite{davoudi2016modeling}. Generally, there are two variants of recommendation approaches: Content based and Collaborative Filtering (CF) based approaches \cite{Adomavicius2011,Ricci2011}. The basic idea of the content based approach is to use properties of an item to predict a user's interests towards it \cite{Yang2014}. The key idea of collaborative filtering is to use the feedback from each individual user \cite{Yang2014}. 

CF approaches can be further grouped into model-based and neighborhood-based \cite{Su2009}. Neighborhood based CF approaches use user-item ratings stored in the system to directly predict ratings for new items \cite{Yang2014}.  In contrast, model-based CF approaches use user-item ratings to learn a predictive model. The general idea is to model the user-item interactions with factors representing latent features of users and items in the system, such as the preference class of users and the category class of items. One of the most accurate approaches was found to be \textit{Matrix Factorization} (MF). The most basic approach to matrix factorization is Singular Value Decomposition (SVD), however numerous more sophisticated approaches have been developed \cite{Koren2008,Fazeli2014,Kabbur2013}.

\cite{Yang2013} and \cite{fang2014leveraging} show that incorporating trust (social relation) into recommender systems has demonstrated potential to improve recommendation performance, and to help mitigate some well-known issues, such as data sparsity and cold start \cite{Guo2012}. A social recommender system improves on the accuracy of the traditional RS by taking social interests and social trusts between users in an social network as additional inputs \cite{Ma2009,liu2010use}. Due to stable and long-lasting social bindings, people are more willing to trust recommendations from their friends more than those from strangers and vendors \cite{Yang2014}. Social trust between a pair of friends may be established based on explicit feedback of user concerning user, or it may be inferred from implicit feedback \cite{Shani2011,Tavakolifard2011}.

To the best of our knowledge, most of the existing social recommendation methods assume that the user preferences may be influenced by a number of explicit social friends \cite{reafee2016power,Ma2011,sun2015recommender}. However, the reliance on social connections may restrict the application of trust-based approaches to other scenarios where social networks are not available or supported \cite{sun2015exploiting}. It is often very challenging to have users giving trust scores of each other. Even these publicly available datasets for trust usually provide trust relations in binary format (0/1), as stated in the literature, because of privacy concerns \cite{Fazeli2014}. In addition, the potential noise and weaker social ties (than trust) in social networks can further hinder the generality of these approaches \cite{sun2015exploiting}. In contrast, in implicit relations social networks, we can only get a user's positive behaviors from the history of what he/she has clicked, purchased or connected \cite{guo2015social}.

Our approach focuses on CF-based social RSs, since collaborative filtering was found to lead to very accurate recommendations in the literature and most existing social recommender systems are CF-based \cite{Koren2008,Jamali2010,Ma2011}. In MF-based social recommendation approaches, user-user social trust information is integrated with user-item feedback history as to improve the accuracy of traditional MF-based RSs, which only factorize user-item feedback data \cite{Yang2014}. Such trust-aware approaches are developed based on the phenomenon that friends often influence each other by recommending items. To investigate this phenomenon, we conduct an empirical trust analysis based on three well-known publicly available datasets (FilmTrust\cite{guo2013novel}, Epinions\cite{konect:massa05}, and Ciao\cite{guo2014etaf}).

In this paper, we build a new recommendation model on top of the state-of-the-art models where both the explicit user-item ratings and implicit social relation involved to improve the accuracy of rating prediction. To the authors' knowledge, our work is the first to extract implicit social relation from ratings-only datasets (trust data is not available) and use in social-based recommendation models. Experimental results on the real-world datasets of recommender systems demonstrate that employing implicit trust into matrix factorization whenever explicit trust is not available, can achieve better accuracy than other counterparts as well as other well-performing recommendation models (ten approaches in total). It should also be noted that in this work, based on the similarity concept, we assume the trust relations are bidirectional and equal in both directions.

The rest of the paper is organized as follows: in Section 2, we summarize the related work and present a brief review of the existing methods. The problem definition and our proposed approach are presented in Section 3. Section 4 presents experimental evaluation of the proposed method with discussions on the results. Finally, we conclude by giving an overview of further work in Section 5. \\

\section{Related Work}
Trust-aware recommender systems have been widely studied, given that social trust provides an alternative view of user preferences other than item ratings \cite{forsati2014matrix}. Yuan et al. \cite{yuan2010improved} found that trust networks are small-world networks where two random users are socially connected in a small distance, indicating the implication of trust in recommender systems. \cite{wang2015social} presented a contextual social network model that takes into account both participants' personal characteristics and mutual relations and proposed a new probabilistic approach, SocialTrust, to social context-aware trust inference in social networks. 

\cite{h2015merging} proposed a multi-view clustering based on Euclidean distance by combination both similarity view and trust relationships that is including explicit and implicit trusts. Hu et al. \cite{hu2016synthetic} proposed a recommendation framework named MR3, which jointly modeled users' rating behaviors, social relationships, and review comments. Liu et al. \cite{liu2016learning} proposed a novel social recommendation method, namely Probabilistic Relational Matrix Factorization (PRMF), which aims to learn the optimal social dependency between users to improve the recommendation accuracy.

There are two main recommendation tasks in recommender systems, namely item recommendation and rating prediction, and our work focuses on the rating prediction task. Matrix factorization technique, because of achieving higher accuracy and better alleviate the data sparsity issue, is a widely-used recommendation method in model-based CF \cite{koren2009matrix}. Trust-aware model-based MF approaches assume that users' preferences are similar to or influenced by their trusted users \cite{guo2016novel}. The intuition behind is that social friends share similar preferences and influence each other by recommending items. It has been shown that such additional side information among users is useful to deal with the concerned issues and thus to improve recommendation performance \cite{sun2015exploiting}.

Specifically, Guo et al. \cite{guo2015leveraging} clustered users by multi-views of similarity and trust, in order to resolve the relative low accuracy. Ma et al. \cite{Ma2008} proposed a social regularization method (SoRec) by considering the constraint of social relationships. Ma, King, and Lyu \cite{Ma2009} proposed a social trust ensemble method (RSTE) to linearly combine a basic matrix factorization model and a trust-based neighborhood model together. They proposed SoReg method that the active user's user-specific vector should be close to the average of her trusted neighbors, and use it as a regularization to form a new matrix factorization model \cite{Ma2011}.

Jamali and Ester \cite{Jamali2010} built a new model (SocialMF) on top of SoRec by reformulating the contributions of trusted users to the formation of the active user's user-specific vector rather than to the predictions of items. Zhu et al. \cite{zhu2011social} proposed a graph Laplacian regularizer to capture the potentially social relationships among users, and form the social recommendation problem as a low-rank semidefinite problem. Zhang et al. proposed a social recommendation method in \cite{zhang2013collaborative}, which the authors utilize as trust network information in the experimental process. Yang et al. \cite{Yang2013} proposed a hybrid method (TrustMF) that combines both a truster model and a trustee model from the perspectives of trusters and trustees. Tang et al. \cite{tang2013exploiting} considered both global and local trust as the contextual information in their model. \cite{yao2014modeling} took into consideration both the explicit and implicit interactions among trusters and trustees in a recommendation model.

Huang et al. \cite{huang2013enhancing} and \cite{sun2015recommender}, only used the social context information, such as tagging and did not incorporate the situation of implicit friendship between users. Fang, Bao, and Zhang \cite{fang2014leveraging} decomposed trust into four general factors and then integrate them into a matrix factorization model. \cite{reafee2016power} focused on the leverage of the hidden social relations between users. Accordingly, they have investigated the power of link prediction techniques to extract the implicit friendship and incorporated it with explicit friendship into probabilistic matrix factorization. Guo et al. \cite{Guo2015} extended SVD++ with social trust information and proposed TrustSVD, a trust-based matrix factorization technique. However, it is also noted that even the latest work \cite{Guo2015} can be inferior to other well-performing ratings-only models. All these works have shown that a matrix factorization model regularized by trust outperforms the one without trust. That is, trust is helpful in improving predictive accuracy. However, there are certain drawbacks among the previous studies. 

In contrast to the incorporation of the explicit friendship relation, there may be implicit correlations between users based on rating matrix. But, the majority of the literature on social recommendation ignores the role of the implicit friendship relation in boosting the accuracy of the recommendations specially whenever explicit trust is not available \cite{reafee2016power}. In this paper, our method differs from the previous work because we present how social relation can be extracted from users' ratings to items by describing Hellinger distance between users in recommender systems and propose to incorporate the predicted trust scores into social matrix factorization models for improving recommendation performance. In this paper, due to better accuracy in rating prediction compared to the similar works, we take Guo et al. \cite{Guo2015} work as a baseline to verify the effectiveness of our method. \\

\section{Proposed Method}
As we know, Recommender System is one of the most important systems that can be modeled by the bipartite networks in which two types of nodes are users and items. In this paper, we want to take user ratings (or any other user behavior) to extract implicit social relation based on what users with similar behavior liked or purchased. The extracted social relation can indicate behavioral similarity between two users in the recommender system.

The first step is measuring the behavioral similarity for each pair of nodes in the bipartite network. We want to choose a proper distance measure as a base metric, because the similarity measures are in some sense the inverse of the distance metrics. We use one type of \textit{f-divergence} metrics \cite{csiszar2004}, called \textbf{Hellinger distance} (also known as \textit{Bhattacharyya distance}), that was introduced by \textit{Ernst Hellinger} in 1909 \cite{Nikulin2001} to quantify the similarity between two probability distributions \cite{VanderVaart1998}. This measure is based on a well-defined mathematical metric \cite{hunter2012}. One of the reasons for choosing Hellinger distance, is the existence of inherent stochastic and statistical properties in this problem, because it is not clearly deterministic. Therefore, we need a statistical metric to measure the distances between the nodes.
The distinctive feature of Hellinger distance compared to other statistical distance is the satisfaction of triangle inequality, that differences between people in a network will be represented properly.

Now, we apply this metric to bipartite networks using degree distribution of neighbors of each node for measuring the similarity of the nodes on one side of the bipartite network. Let $L_i$ be the number of $x$'s neighbors with degree of $i$. Suppose the vector $L_x=(l_1,\dots,l_{\Delta})$ be the non-normalized distribution of $L_i$ for all adjacent neighbors of $x$. We introduce the Hellinger distance between two nodes, $x$ and $y$, on one side of the bipartite network as follows: 
\begin{eqnarray}
d(x,y) = \sqrt{2}\ D_H (L_x\|L_y)
\end{eqnarray}
where the f\textnormal{-}divergence of $L_x$ from $L_y$ measures the difference between two probability distributions $L_x$ and $L_y$, and the function $d(x,y)$ represents the normalized form of this difference. 

Now, we generate an $N\times N$ distance matrix ($N$ is the number of nodes in users side of bipartite network). According to the well-defined metric features and the ability of mapping to Euclidean space, we can form social relations based on how close nodes are to each other. It means that any pair of nodes in the matrix with a less distance can be formed one tie (i.e. link, edge) by specific neighborhood radius. So, we can consider ties between nodes as the Hellinger distance between two nodes is lesser than certain threshold and make a new social network between users. But taking into account an appropriate threshold is extremely important to address in the next subsection. 

Calculation of non-normalized degrees distribution of nodes and update nodes vectors on two sides of each link (degree of nodes), both takes $O(m)$ time, where $m$ is the number of links in the graph. Also, finding Hellinger distance of the nodes takes $O(n^2\Delta)$ time, that $n$ is the number of all nodes in one part of the graph and $\Delta$ is the length of the $L_x$ vector. Then in overall, obtaining Hellinger distance matrix requires $O(m+n^2\Delta)$ time. 

As a result, we have edge list of implicit social relations between users part of recommender systems and we can employ this edge list into matrix factorization social recommendation techniques for evaluation of our extracted social relation in user rating prediction. \\

\subsection{Find Desired Threshold}
Finding threshold is an important challenge in our problem, and here we use an ad-hoc or heuristic approach. To earn the required global threshold, the likelihood of distance between different users on the network can be used. We assume that the distribution of the two users' distance can be modeled in terms of set of unknown parameters $\theta$. Suppose that $d_1,d_2,\dots,d_n$ are independent random selection of the elements in the obtained distance matrix. If we want to find the threshold for each individual user on the network, we can do random selection of the distances only on same row of the user; in this work, we supposed distances are independent of each other and we choose $d_i$'s randomly from the whole network.  $P(D|\theta)$ is the probability density of D (distances), given the distribution parameter $\theta$ which is defined as:
\begin{equation}
P(D|\theta) = \prod_i P(d_i,|\theta)
\end{equation}

In fact, to find threshold, we use the estimated distance distribution of nodes by sampling from main graph. This means that by having information of the neighbors of nodes and samples of the paired nodes in entire graph, we can obtain an estimate of this distribution. We assume the distances between nodes are coming from a Normal distribution, where $\mu$ and $\sigma^2$ are specific parameters of this distribution function, then the maximum likelihood estimators of the mean and the variance of the normal distribution will be \cite{harris1998handbook}:
\begin{eqnarray}
\widehat{\mu}_n &=& \frac{1}{n} \sum^n_{j-1} d_j \\
\widehat{\sigma}^2_n &=& \frac{1}{n} \sum^n_{j-1}(d_j - \widehat{\mu})^2 
\end{eqnarray}
Thus, the estimator $\widehat{\mu}$ is equal to the sample mean and the estimator $\widehat{\sigma}^2$ is equal to the unadjusted sample variance.

Of course, this distribution is not necessarily normal, and normal is just one example to illustrate the point. Accordingly, our desired threshold value will be dependent to the density of the social graph of users to extract social information. It means that the distinction between sparse and dense social graphs is rather vague, and depends on the context. A dense graph is a graph in which the number of edges is close to the maximal number of edges. The opposite, a graph with only a few edges, is a sparse graph. For finding threshold in the bipartite graph, we define the graph density, $\alpha$ as follows:
\begin{equation}
\alpha = \frac{E[deg]}{M}
\end{equation}
where $E[deg]$ is the desirable variable of expected value of degrees of social graph nodes and $M$ is the number of nodes in the other side of bipartite graph (the maximum number of edges for each user node). So, to determine the threshold for extracting social edges, we convert normal
distribution of distances between nodes to a standard normal and find the corresponding probability. It means that for the threshold $T$, we have:

\begin{equation}
\Phi(d) = P(d_i \leq T) = \alpha
\end{equation}
where $\Phi(d)$ is the normal distribution function which gives the probability that a standard normal variate assumes a value in the interval $[0,d]$.

Now, we can use inverse of cumulative distribution function of the standard normal distribution, $\Phi^{-1}(d)$, for finding our favorite threshold (depending on the amount of density in the extracted social graph). Therefore,  we can determine the threshold using the information from the network graph (without trust information). As a result, after finding the own desired threshold value, if the distance value are less than the threshold amount, friendship and trust edge can formed in new social network by using the distances for all users. \\

\subsection{Matrix Factorization: A Basic Model}
In general, in recommender systems, we have a set of users $\{u_1, \dots, u_N\}$ and set of items $\{i_1 , \dots, i_M\}$. The rating matrix $R = [R_{u,i} ]_{N\times M}$ provide the ratings given by users to items. Therefore, $R_{u,i}$ is the rating of user $u$ to item $i$. The recommender system's task is then to predict rating of user $u$ to item $i$ whenever $R_{u,i}$ is unknown. For clarity, we preserve symbols $u, v$ for users, and $i, j$ for items. Let $I_u$ denote the set of items rated by user $u$.

Rating scores are the explicit user feedback and Matrix Factorization (MF) is a state-of-the-art recommender method to exploit this rating information \cite{hu2016synthetic}. MF techniques have gained popularity and become the standard recommender approaches due to their accuracy and scalability \cite{koren2009matrix}. The goal of matrix factorization is to learn latent features, and subsequently, to employ them for making rating predictions \cite{Koren2008}. Let $p_u$ and $q_i$ be a $L$-dimensional latent feature vectors of user $u$ and item $i$, respectively. The essence of matrix factorization is to find two low-rank matrices: user-feature matrix $P \in \mathbf{R}^{L\times N}$ and item-feature matrix $Q \in \mathbf{R}^{L\times M}$ that can adequately recover the rating matrix $R$, i.e., $R \approx P^T Q$, where $P^T$ is the transpose of matrix $P$. Hence, the rating on item $j$ for user $u$ can be predicted by the inner product of user-specific vector $p_u$ to item-specific vector $q_j$, i.e., $\hat{r}_{u,j} = q^T_j p_u $. In this regard, the main task of recommendations is to predict the rating $\hat{r}_{u,j}$ as close as possible to the ground truth $r_{u,j}$. Formally, we can learn the user- and item-feature matrices by minimizing the following loss (objective) function \cite{guo2016novel}:

\begin{eqnarray}
L_r = \frac{1}{2}\sum_u \sum_{j \in I{u}} (\hat{r}_{u,j}-r_{u,j})^2 +\frac{\lambda}{2} (||P||^2_F+||Q||^2_D) 
\end{eqnarray}

and the predicted ratings is:
\begin{eqnarray}
\hat{r}_{u,j} = \mu + b_u + b_j + p_{u}^T q_{j} 
\end{eqnarray}
where $\Vert . \Vert_F$ denotes the Frobenius norm, and regularization parameter $\lambda$ controls over-fitting. The rating mean is captured by $\mu$; $b_u$ and $b_j$ are rating biases of $p_u$ and of $q_j$. The $L$-dimensional feature vectors $p_u$ and $q_j$ represent preferences for user $u$ and characteristics for item $j$, respectively. The dot products $p_{u}^T q_{j}$ capture the interaction or match degree between users and items \cite{hu2016synthetic}. \\

\subsection{Social Recommender System}
In this paper, we follow the basic idea of matrix factorization method in social recommender systems to learn the latent features of both users and items more precisely when trust information between users is not available. 

Now, based on proposed method in \cite{Guo2015}, it is assumed that $T_{u,v}$ denotes the trust value between users $u$ and $v$. Therefore, matrix $T = [T_{u,v}]_{N\times N}$ represent all trust scores between users, where $T_u$ is the set of users trusted by user $u$. Note that $T$ can be asymmetric in general. $p_u$ and $w_v$ are denoted as the $L$-dimensional latent feature vectors of truster $u$ and trustee $v$, respectively. The trusters and the active users in the rating matrix in the trust matrix are limited to share the same user-feature space. Hence, we have truster-feature matrix $P^{L\times N}$ and trustee-feature matrix $W^{L\times N}$. By employing the low-rank matrix approximation, the trust matrix can be recovered by $T \approx P^T W$. Thus, a trust relationship can be predicted by the inner product of a truster-specific vector and a trustee- specific vector $\hat{t}_{u,v} = w^T_v p_u$ \cite{guo2016novel}.

Our model, called Hell-TrustSVD, is built on top of the state-of-the-art model known as TrustSVD proposed by Guo et al. \cite{Guo2015}. The Hell-TrustSVD model exploits trust scores to learn the latent features more precisely. The rationale behind TrustSVD is adopt a distinct strategy that the popular users and items should be less penalized, and cold-start users and niche items (those receiving few ratings) should be more regularized. Therefore, term $T_v^+$ in the loss function of equation (1) in paper \cite{Guo2015} can be replace by $(Hell\textnormal{-}T)_v^+$, which is the set of users who trust user $v$ that extracted by Hellinger distance as we mentioned before. \\ 

\section{Experimental Evaluation}
\subsection{Dataset Description}
The experiments are performed on five real-world datasets:
 FilmTrust\cite{guo2013novel}, Epinions\cite{konect:massa05}, Ciao\cite{guo2014etaf}, MovieLens-100K\cite{harper2016movielens}, and MovieLens-1M\cite{harper2016movielens}. FilmTrust consists of 35,497 ratings given by 1,508 users to 2,071 movies. On Ciao dataset, we have 280,391 ratings given by 7,375 users to 99,746 items. On Epinions dataset, there are 664,824 ratings given by 40,163 users to 139,738 items. MovieLens-100K contains 100,000 ratings given by 943 users to 1,682 movies and MovieLens-1M includes 1,00,000 ratings given by 6,040 users to 3,706 movies. The densities of the rating matrices on these datasets are 6.30\% for MovieLens-100K, 4.47\% for MovieLens-1M, 1.14\% for FilmTrust, 0.03\% for Ciao, and 0.051\% for Epinions. Moreover, on FilmTrust, Ciao and Epinions datasets, we have observed 1,853 , 111,781 and 487,183 social relationships between users. The densities of the social relation matrices are 0.042\% for FilmTrust, 0.23\% for Ciao and 0.029\% for Epinions.

For each dataset, we use 5-fold cross-validation for learning and testing. Specifically, we randomly split each data set into five folds and in each iteration four folds are used as the training set and the remaining fold as the test set. Five iterations will be conducted to ensure that all folds are tested. The average test performance is given as the final result. \\

\subsection{Evaluation Metric}
The performances of the recommendation algorithms are evaluated by two most popular metrics: Mean Absolute Error (MAE) and Root Mean Square Error (RMSE). The definitions of MAE and RMSE are as follows:
\begin{equation}
MAE = \frac {\sum_{u,i}{|\hat{r}_{u,i}-r_{u,i}|}} {N_{test}} 
\end{equation}
\begin{equation}
RMSE = \sqrt{\frac {\sum_{u,i}{(\hat{r}_{u,i}-r_{u,i})^2}} {N_{test}}} 
\end{equation}
where $N_{test}$ is the number of test ratings, $r_{u,i}$ denotes the observed rating in the testing data, and $\hat{r}_{u,i}$ is the predicted rating. Smaller values of MAE and RMSE indicate better predictive accuracy. \\

\subsection{Comparing Method}
Two kinds of approaches are compared with our proposed method (Hell-TrustSVD): (1) Baselines (only ratings are available) models:  GlobalAvg, UserAvg, ItemAvg, SlopeOne, UserKNN, ItemKNN, RegSVD, BiasedMF and\ SVD++ \cite{Koren2008}; (2) Trust-based models: SoRec \cite{Ma2008}, SoReg \cite{Ma2011}, TrustMF \cite{Yang2013} and TrustSVD \cite{Guo2015}. \\

\subsection{Parameter Setting}
The optimal experimental settings for each method are determined either by our experiments or suggested by previous works. The settings are: (1) SoRec: the number of latent features $d = 5$ for SoRec-1 and $d=10$ for SoRec-2, and $\lambda_c= 1.0, 0.01,1.0$ corresponding to FilmTrust, CiaoDVD and Epinions respectively; (2) SoReg: the number of latent features $d = 5$ for SoReg-1 and $d=10$ for SoReg-2, and $\beta = 0.1$ for the all; (3) TrustMF: : the number of latent features $d = 5$  and $\lambda_t = 1$ for all and Tr, Te and T (resp.); (4) TrustSVD: the number of latent features $d = 5$ for TrustSVD-1 and $d=10$ for TrustSVD-2, and $\lambda = 0.1, \lambda_t = 0.9$ for FilmTrust, $\lambda = 0.6, \lambda_t = 0.5$  for Epinions, and$\lambda = 0.5, \lambda_t = 1.0$ for CiaoDVD.

Specifically, the optimal experimental settings for baselines and the ratings-only state-of-the-art models is shown in tables \ref{table:comp_socials} and \ref{table:soc_rec_comp_ml}.

\subsection{Comparison with Other Models}
In first experiment, we want to examine accuracy of extracted social relations compared to explicit social relations and trust information. Thus, three recommender datasets that their trust and social information between users are available, has been considered, and by using some of the most popular social recommender models, the comparison between the actual and extractive trust data sets has been done.

 \begin{table*}[!ht]
 \centering
\renewcommand{\arraystretch}{1.3}
\caption{\small Performance comparison details in real based and extracted based social relations in social recommendation models (Hellinger based outputs are shown in parenthesis)}
\label{table:comp_socials}
\begin{small}
\begin{tabular}{|c|c|c|c|c|c|c|}
\hline
\multirow{2}{*}{\textbf{Algorithm}} & \multicolumn{2}{c|}{\textbf{FilmTrust}} & \multicolumn{2}{c|}{\textbf{CiaoDVD}}& \multicolumn{2}{c|}{\textbf{Epinions}} \\
\cline{2-7}
 &  \textbf{MAE} & \textbf{RMSE} & \textbf{MAE} & \textbf{RMSE} & \textbf{MAE} & \textbf{RMSE}\\
\hline
SoReg1 & 0.64	(0.644)&0.829(0.831)&0.903(0.906)	&1.169(1.17) & 0.963	(0.963)	& 1.263(1.263) \\
 \hline
 SoReg2 &0.682(0.684)&0.885(0.886)&0.771(0.772)&1.044(1.048)& 0.914(0.913)&1.205(1.204)\\
  \hline
SoRec1 & 0.728(0.768)	&0.937(0.974)&0.849(0.851)&1.056(1.058)& 0.934(0.973)	&1.17(1.218)\\
  \hline
  SoRec-2 & 0.728(0.766) &0.906(0.942)&0.813(0.814)&1.105(1.104) & 0.92(0.957) &	1.241(1.258) \\
  \hline
  TrustSVD-1 & 0.623(0.625) &	0.801(0.804) &0.72(0.793)&0.936(1.045)& 0.82(0.827)& 1.051(1.063)  \\
  \hline
  TrustSVD-2 & 0.61(0.621) &	0.79(0.799) &0.721(0.798)&0.938(1.051)& 0.817(0.826) & 1.048(1.064)  \\
  \hline
  TrustMF-Tr & 0.63(0.631)& 0.813(0.808)&0.767(0.779)	&0.995(1.017)&0.833(0.852) & 1.086 (1.086) \\
  \hline
  TrustMF-Te &0.631(0.631) &0.814(0.809) &0.767(0.766)&0.995(1.001)& 0.832(0.85) &	1.085(1.085) \\
  \hline
  TrustMF-T &0.628(0.638	)&0.808(0.814)&0.765(0.764)&0.988(0.992)& 0.834(0.859) &	1.089(1.09) \\
  \hline
\end{tabular}
\end{small}
\end{table*}

\begin{figure*}[!ht]
\centering
\subfloat[FilmTrust Compare in term of MAE]{\includegraphics[width=2.3in]{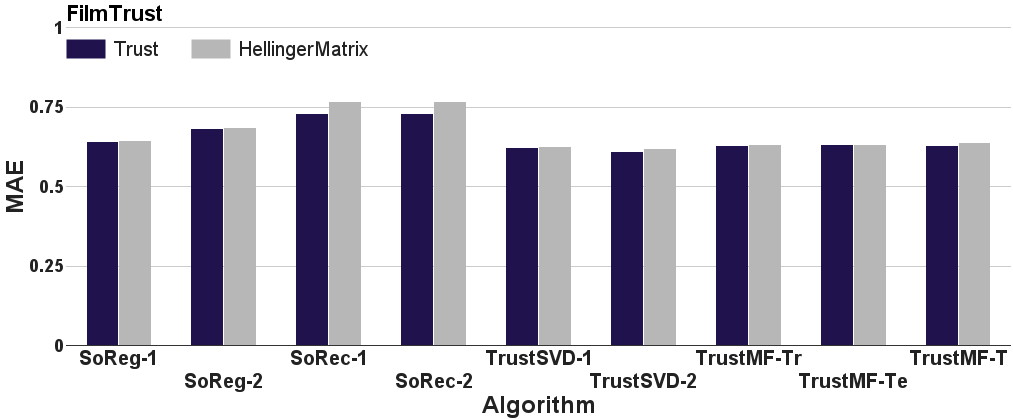}\label{FilmTrustCompare_MAE}}
\hfil
\subfloat[CiaoDVD Compare in term of MAE]{\includegraphics[width=2.3in]{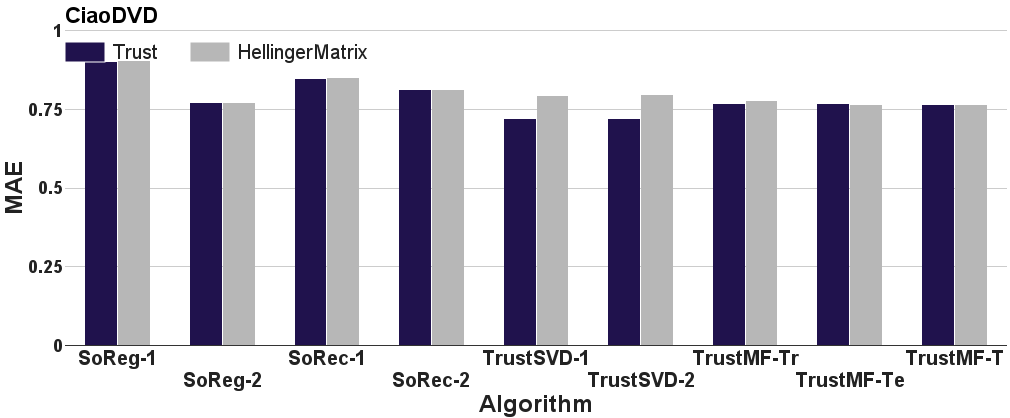}\label{CiaoDVDCompare_MAE}}
\hfil
\subfloat[Epininos Compare in term of MAE]{\includegraphics[width=2.3in]{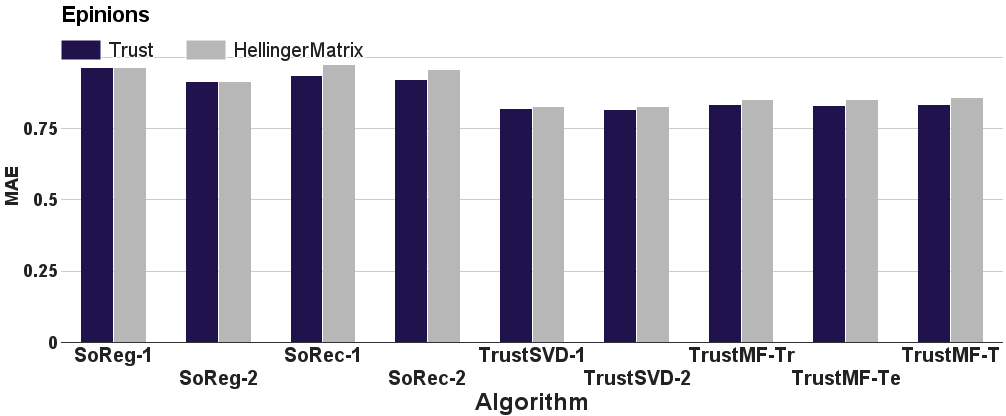}\label{EpininosCompare_MAE}}
\hfil
\subfloat[FilmTrust Compare in term of RMSE ]{\includegraphics[width=2.3in]{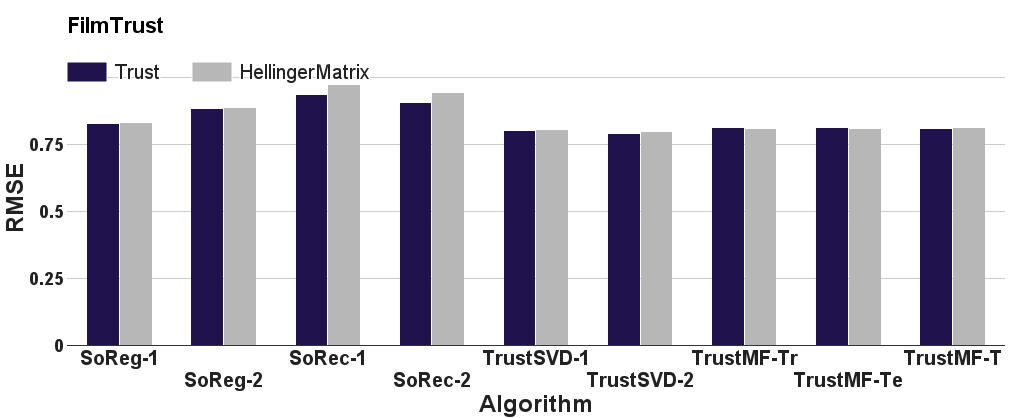}\label{FilmTrustCompare_RMSE}}
\hfil
\subfloat[CiaoDVD Compare in term of RMSE]{\includegraphics[width=2.3in]{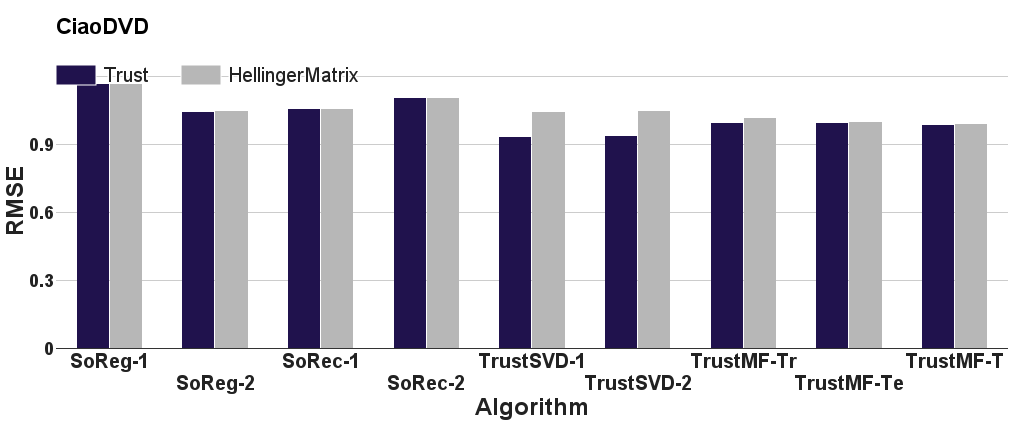}\label{CiaoDVDCompare_RMSE}}
\hfil
\subfloat[Epininos Compare in term of RMSE]{\includegraphics[width=2.3in]{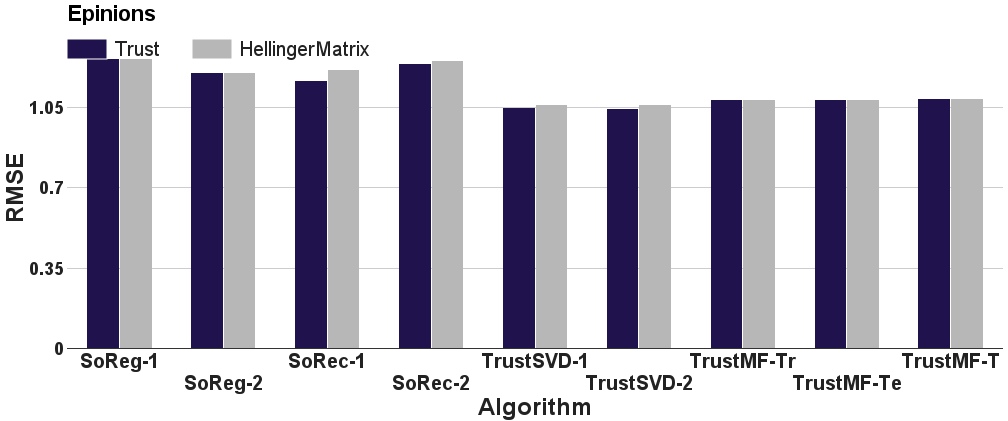}\label{EpininosCompare_RMSE}}
\caption{\small Comparison between real based and extracted based social relations social recommendation models}
\label{subfig_comp_socials}
\end{figure*}

Table \ref{table:comp_socials} and Figure \ref{subfig_comp_socials} shows that social matrix factorization models based on extracted social data are quite similar to the results of models using explicit trust for each series of social recommender datasets. As the Figure shows, the models on implicit trust inferred Hellinger distance can perform as accurate as the models with explicit trust. Regarding our research question in Section 1, we may safely conclude that the implicit trust can be incorporated into the social matrix factorization whenever explicit trust is not available. Moreover, the results in Table \ref{table:comp_socials} and Figure \ref{subfig_comp_socials} conform to the results that TrustSVD was selected as the best candidate for inferring trust scores. However, there is no insistence to use this model. If there is any other better social recommendation models, our method we will be able to use it. Therefor, it indicates that social information obtained from Hellinger distance method for estimating similarity of users behavior, has a good performance in social recommender models. 

\begin{figure}[!ht]
\centering
    \includegraphics[width=0.5\textwidth]{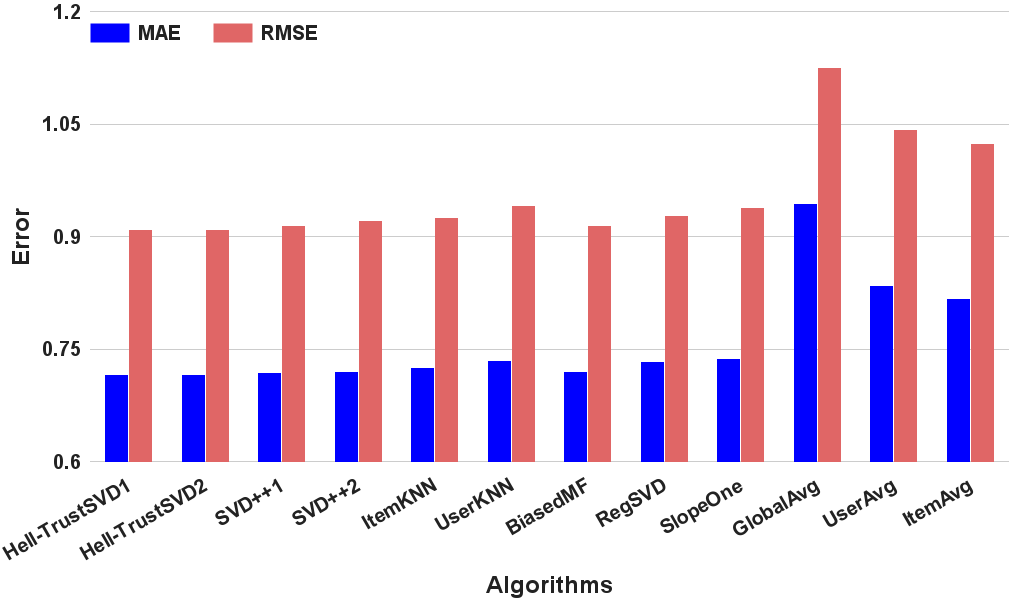}
    \caption{\small MAE and RMSE in MovieLens (100K)}
    \label{fig:ml_100k_compares}
\end{figure}

\begin{figure}[!ht]
\centering
    \includegraphics[width=0.5\textwidth]{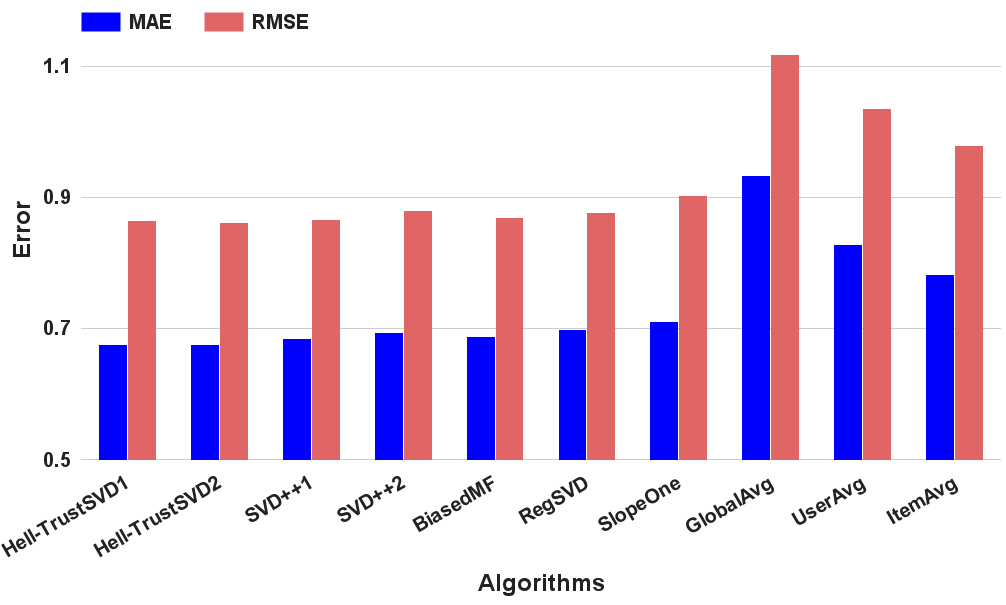}
    \caption{\small MAE and RMSE in MovieLens (1M)}
    \label{fig:ml_1m_compares}
\end{figure}

Now for evaluating proposed approach for recommendations data sets that social data are not available and there are only users' ratings, we do an experiment for these categories. The experimental results are presented in Figure \ref{fig:ml_100k_compares} and Figure \ref{fig:ml_1m_compares} and also details of evaluated models configurations are inserted in Table \ref{table:soc_rec_comp_ml}. For all the comparing methods in the testing view of All, Hell-TrustSVD outperforms the other methods in Movielens-100k and Movielens-1M. Although the percentage of relative improvements are small, Koren \cite{koren2010factor} has pointed out that even small improvements in MAE and RMSE may lead to significant differences of recommendations in practice.

One more observation from Table \ref{table:soc_rec_comp_ml} that the performance of Hell-TrustSVD when d = 5 is very close to that when d = 10, indicating the reliability of our approach with respect to the feature dimensionality. We ascribe this feature to the consideration of both the explicit and implicit influence of ratings and trust in a unified recommendation model.

In conclusion, the experimental results indicate that our approach Hell-TrustSVD outperforms the other methods in predicting more accurate ratings in only ratings recommender dataset, and that its performance is reliable with different number of latent features. As shown in Table \ref{table:comp_socials} and Figure \ref{subfig_comp_socials}, this method of extraction social relationships, has the ability of adding to all kinds of social recommender model as well.

 \begin{table*}[!ht]
 \centering
\renewcommand{\arraystretch}{1.3}
\caption{\small Performance comparison details in MovieLens (100K) and (1M) datasets (standard errors reported)}
\label{table:soc_rec_comp_ml}
\begin{small}
\begin{tabular}{|c|c|c|c|c|c|}
\hline
\multirow{2}{*}{\textbf{Algorithm}} & \multicolumn{2}{c|}{\textbf{MovieLens (100K)}} & \multicolumn{2}{c|}{\textbf{MovieLens (1M)}}& \multirow{2}{*}{\textbf{Configuration}} \\
\cline{2-5}
 & \textbf{MAE} & \textbf{RMSE} & \textbf{MAE} & \textbf{RMSE} & \\
\hline
GlobalAvg & 0.944(0.001)	& 1.125(0.001) & 0.933(0.001)	& 1.117(0.001) &\\
 \hline
 UserAvg &0.835(0.001)&1.042(0.001)& 0.828(0.001)&1.035(0.001)&\\
  \hline
  ItemAvg & 0.817(0.001)	&1.024(0.001)& 0.782(0.001)	&0.979(0.001)&\\
  \hline
  SlopeOne  & 0.738(0.001) &0.939(0.001) & 0.711(0.001) &	0.902(0.001) &\\
  \hline
  UserKNN & 0.735(0.002) &	0.942(0.002) & 0.705(0.002)& 0.907(0.002) & similarity=PCC, shrinkage=30, neighbors=50\\
  \hline
  ItemKNN & 0.725(0.002) &	0.925(0.001) & 0.69(0.002) & 0.877(0.002) & similarity=PCC, shrinkage=30, neighbors=50\\
  \hline
  RegSVD & 0.733(0.003)& 0.928 (0.008)&  0.698	(0.02) & 0.877 (0.02)& factors=10, max.iter=200,learn.rate=0.01,reg=0.1\\
  \hline
  BiasedMF & 0.72(0.002) &	0.914(0.004) & 0.688(0.01) &	0.869(0.01) &  factors=10,max.iter=200,learn.rate=0.01, reg=0.1 \\
  \hline
  SVD++ & 0.72 (0.002)&	0.921 (0.01)& 0.694(0.02) &	0.879 (0.02)& factors=5,max.iter=100,learn.rate=0.01 ,reg=0.1\\
  \hline
  SVD++ & 0.719(0.001) &	0.914(0.003) &  0.685(0.02) &	0.866(0.01) & factors=10,max.iter=100,learn.rate=0.01 ,reg=0.1\\
  \hline
 HellTrustSVD  & 0.716(0.001) &	0.909(0.002) & 0.675(0.01) &	0.861(0.02) & factors=5,max.iter=200,learn.rate=0.001, reg=0.1\\
 \hline
 HellTrustSVD  & 0.716(0.002) &	0.909(0.001) & 0.676(0.01) &	0.865(0.02) & factors=10,max.iter=200,learn.rate=0.001, reg=0.1\\
 \hline
\end{tabular}
\end{small}
\end{table*}

\subsection{Impact of Social Threshold $T$ }
\begin{figure}[!ht]
\centering
\subfloat[FilmTrust Compare MAE]{\includegraphics[width=3.45in]{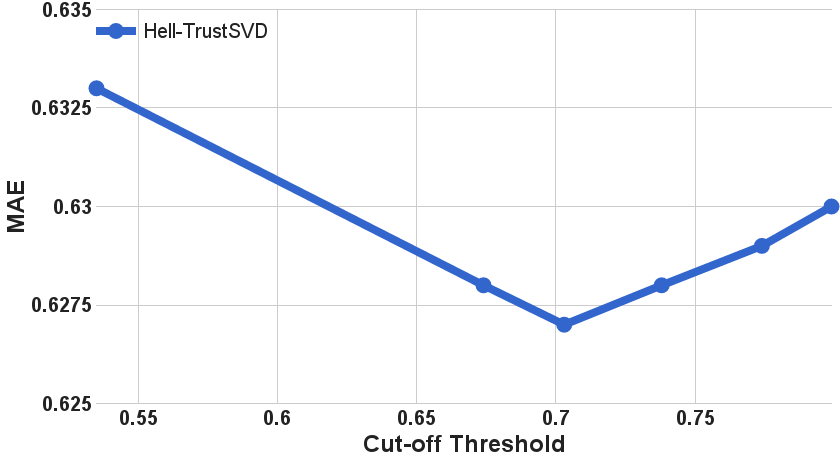}}
\hfil
\subfloat[FilmTrust Compare RMSE ]{\includegraphics[width=3.45in]{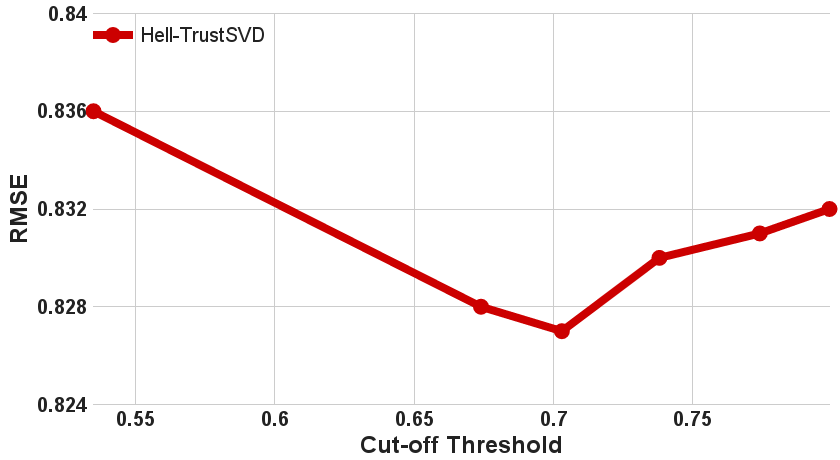}}
\caption{\small Variations of MAE and RMSE compared to different T in the FilmTrust (num.factors=5, max.iter=50, learn.rate=0.005, $\lambda$=0.5)
}
\label{fig:FilmTrust_diff_cut-off}
\end{figure}

As described in Section 3.1, our model has a threshold $T$, which this amount is dependent on parameter $E[deg]$ that indicates the impact of social graph density in the extracting social trust edge. To analyze how sensitive our model is to this parameter, Figure \ref{fig:FilmTrust_diff_cut-off} shows the effect of changing the value of $E[deg]$ on the MAE and RMSE in the range of [0.1, 500] on FilmTrust dataset. For example, if the expected degree of our desired social network is equal to 1, as regards the number of movies in the FilmTrust ($M=2071$) and $\mu=0.801$ and $\sigma^2=0.038$, the threshold T will be 0.674. Specifically, we show the MAE and RMSE of the Hell-TrustSVD model for different values of this parameter. As shown in this figure, value of\ 0.703\ for $T$ (10 for expected of degrees) seems to be a good candidate for our experiment since the Hell-TrustSVD model provides the lowest MAE and RMSE at this value.

\section{Conclusion and Future Work}
We presented a novel trust-based recommendation approach, called Hell-TrustSVD, to mitigate the research gap between user similarity and trust concepts in recommender systems. Specifically, we first extracted the behavioral representation between users based only on ratings data and explore social edges. Then, we incorporated these inferred trust scores into a matrix factorization method in social recommender system. Indeed, we combine a user's influence to other users trusting the user, and that to other users trusted by the user, where their trust information is utilized to update the user latent feature vectors. We conducted comparisons with the state-of-the-art approaches. The results show that the Hell-TrustSVD with implicit trust performs in ways similar to the TrustSVD using explicit trust. A clear advantage of this result is that, since we often have no trust scores explicitly given by users in social networks, we can overcome this problem by using implicit (or inferred) trust scores and incorporate them into the recommender.

For future work, we aim to define and infer trust scores taking into account context data of users rather than their ratings only and extend our model by considering other properties (i.e. trust weight) of trust networks to further improve recommendation accuracy. We also want to evaluate additional dimensions of recommendation quality, such as diversity, novelty or serendipity and improve the proposed model by considering both the influence of trusters and trustees. Furthermore, we want to use compressive sensing in networks \cite{mahyar2013ucsnt,mahyar2013ucswn,Mahyar2015CScomdet,Mahyar2015LSRweighted} to efficiently identify top-$k$ influential users \cite{Mahyar2015TopK} via extracted social relations.

%
\bibliographystyle{abbrv}
\bibliography{references} 


\end{document}